\documentclass[doublespacing]{elsart_no_numbers}
\usepackage{natbib}
\usepackage{graphics}
\usepackage{graphicx}
\usepackage{epsfig}
\usepackage{amssymb}
\usepackage{amsfonts}
\usepackage{amsmath}

\begin{document}

\begin{frontmatter}

\title{Pseudo-potentials and loading surfaces for an endochronic
plasticity theory with isotropic damage}

\author{Silvano Erlicher \corauthref{cor1}},
\author{Nelly Point \corauthref{aut2}}

\corauth[cor1]{Researcher, Universit\'{e} Paris-Est, UR Navier,
Ecole des Ponts, LAMI, \\ 6 et 8 av. B. Pascal, Cit\'{e} Descartes,
Champs-sur-Marne, 77455 Marne-la-Vall\'{e}e, \\  Cedex 2, France
(corresponding author). E-mail: erlicher@lami.enpc.fr \\}

\corauth[aut2]{Professor, Conservatoire National des Arts et
M\'{e}tiers (CNAM), Sp\'{e}cialit\'{e} \\ Math\'{e}matiques (442),
292 rue Saint-Martin, 75141 Paris, Cedex 03, France. \\ E-mail:
point@cnam.fr}

\begin{abstract}
The endochronic theory, developed in the early seventies, allows the
plastic behavior of materials to be represented by introducing the
notion of intrinsic time. With different viewpoints, several authors
discussed the relationship between this theory and the classical
theory of plasticity. Two major differences are the presence of
plastic strains during unloading phases and the absence of an
elastic domain. Later, the endochronic plasticity theory was
modified in order to introduce the effect of damage. In the present
paper, a basic endochronic model with isotropic damage is formulated
starting from the postulate of strain equivalence. Unlike the
previous similar analyses, in this presentation the formal tools
chosen to formulate the model are those of convex analysis, often
used in classical plasticity: namely pseudo-potentials, indicator
functions, sub-differentials, etc. As a result, the notion of
\emph{loading surface} for an endochronic model of plasticity with
damage is investigated and an insightful comparison with classical
models is made possible. A damage pseudo-potential definition
allowing a very general damage evolution is given.

CE DATABASE SUBJECT HEADINGS: Plasticity, Thermodynamics, Damage,
Constitutive models

\end{abstract}

\end{frontmatter}

\clearpage \setlength{\hoffset}{-1cm}

\section{Introduction}

In the early seventies, \citet{Valanis71} proposed the \emph{endochronic
theory of visco-plasticity}, which postulates the existence of an \emph{%
intrinsic time} governing the rate-independent evolution of stress and
strains in materials, whereas the Newtonian time is exploited to model the
viscous behavior; see also \citep{Schapery68,Bazant76}. In the case of
plasticity without viscous effects, the resulting constitutive laws are
characterized by the absence of an elastic domain and the corresponding
hysteresis loops are typically smooth and open. The flow rules of these
models were not originally formulated in terms of pseudo-potentials, which
made the direct comparison of this class of models with classical plasticity
theories difficult \citep{Valanis80}. However, it was recently proven by %
\citet{ErlicherPoint2006} that endochronic models do admit a representation
based on pseudo-potentials and on the normality assumption, provided that
pseudo-potentials be endowed with an additional dependence on state
variables. This proof, given for the case of plastically incompressible
models, showed the strong relationship between the endochronic theory and
the generalized plasticity %
\citep{Phillips65,Eisenberg71,Lubliner93,Auricchio95}. It was also shown
that the non-linear kinematic hardening model, that is associated, but is
not in a generalized sense, admits a representation in terms of a
pseudo-potential. Recently, the same authors extended this analysis to other
models, like the Mr\'{o}z model \citep{PointErlicher2007} and the
non-associated Drucker-Prager model \citep{ErlicherPoint2005}; see also %
\citet{Ziegler87}, \citet{Houlsby2000}. In summary, this thermodynamically
well-posed approach can be used for a very large class of existing \emph{%
classical} or \emph{non-classical} plasticity models. Actually, a similar
approach is used in geotechnical engineering, see e.g. %
\citet{CollinsHoulsby1997}, where pseudo-potentials have an additional
dependence on the so-called \emph{true stresses}, distinguished from the
\emph{generalized stresses}.

The standard endochronic theory was modified by several authors through the
introduction of a damage variable. Using the strain equivalence postulate, %
\citet{Xiaode1989} proposed an endochronic model with isotropic damage,
while \citet{Valanis1990} discussed an endochronic model with anisotropic
damage, in the larger theoretical framework of fracture mechanics. Later, a
different approach based on the postulate of energy equivalence was used,
among others, by \citet{ChowChen1992} and %
\citet{WuNanakorn1998,WuNanakorn1999}.

In the aforementioned works, the thermodynamic formulation of flow rules is
not based on the notions of pseudo-potentials and loading surfaces, as it is
typical for other classical plasticity models with or without damage. Hence,
in this paper, a simple endochronic model of plasticity with isotropic
damage similar to that discussed by \citet{Xiaode1989} is presented: no
generalization is introduced with respect to the previously cited models,
but a new approach is suggested for their description. In detail, the
postulate of strain equivalence is adopted; the Helmholtz energy is assumed
to have a regular quadratic term and an additional singular term; the tools
of the convex analysis such as indicator functions and sub-differentials %
\citep{Rockafellar69, Moreau70, Fremond2002} are used to define the flow
rules starting from well-suited pseudo-potentials. This presentation leads
to the proper definition of the plasticity loading surface for an
endochronic model with damage and is a direct extension of the results
concerning the endochronic model without damage already discussed in %
\citet{ErlicherPoint2006}. Only plastically incompressible models are
considered here, since they permit to explain the main ideas, without
introducing a too complex formalism. The extension to the general case is
possible, but it is omitted for simplicity. The proposed analysis has an
intrinsic interest, since it allows an easier comparison between endochronic
models with damage and classical plasticity models with damage. Nonetheless,
in the authors' opinion, another important reason justifies the interest
towards this class of models: they represent the suitable theoretical basis
for the analysis of the thermodynamic admissibility of the Bouc-Wen models
with strength and stiffness degradation; see among others %
\citep{Bouc71,Wen76,Baber81,Casciati89,Karray89}. This was one of the main
motivation at the origin of the present study and the related developments
about degrading Bouc-Wen models are presented in a companion paper %
\citep{ErlicherBursi2007}.

After the introduction, the endochronic theory is presented in the second
section: in the first part, standard endochronic models are described, while
the second part concerns the definition of the flow rules of the extended
endochronic theory, characterized by an additional scalar variable endowed
with damage. The thermodynamic framework, with the definition of the suited
pseudo-potentials, is discussed in the following section and is supplemented
by numerical examples. Then, a brief discussion about stability and
uniqueness is made and the concluding remarks are given, where the topics
dealt with in the companion paper \citep{ErlicherBursi2007} are pointed out.

\section{Endochronic models}

\label{SecEMs}

\subsection{Flow rules of plastically incompressible ND-EC models}

The endochronic theory was first formulated by \citet{Valanis71}, who
suggested the use of a positive scalar variable $\vartheta $, called the
\emph{intrinsic time scale}, in the definition of constitutive plasticity
models. The evolution laws are described by convolution integrals involving
past values of the strain $\mbox{\boldmath $\varepsilon$}$ and suitable
scalar functions $\mu $ depending on $\vartheta $, called \emph{memory
kernels}. When the memory kernel is exponential, the integral expressions
can be rewritten as simple differential equations, the flow rules; in the
case of an isotropic endochronic model without hardening or softening,
called here ND-EC model (see Figure \ref{scheme_Endo}), fulfilling the
plastic incompressibility assumption, they read:
\begin{equation}
\left\{
\begin{array}{l}
tr\left( \mbox{\boldmath $\sigma$}\right) =3K\ tr\left(
\mbox{\boldmath $\varepsilon$}\right) ,\text{ \ \ \ \ \ }dev\left(
\mbox{\boldmath $\sigma$}
\right) =\mathbf{z} \\
\mathbf{\dot{z}}=2G\ dev\left( \mbox{\boldmath
$\dot{\varepsilon}$}\right) \mathbf{-}\beta \
\mathbf{z}\dot{\vartheta}
\end{array}
\right.  \label{EndoFormulGen}
\end{equation}
where $\beta >0$ (notice that $\beta$ different from zero is needed
to have a non elastic behavior); the superposed dot indicates the
time derivative; $ \mbox{\boldmath $\varepsilon$}$ is the small
strain tensor; $\mbox{\boldmath $\sigma$}$ is the Cauchy stress
tensor; $tr$ and $dev$ are the trace and deviatoric operators; $K$
is the bulk modulus while $G$ is the shear modulus. The simplest
choice for the \emph{intrinsic time scale} flow indicated in
(\ref{EndoFormulGen}) is $\dot{\vartheta}=\left\Vert dev\left(
\mbox{\boldmath $\dot{\varepsilon}$}\right) \right\Vert $. It is
interesting to note that relationships (\ref{EndoFormulGen}) are
equivalent to
\begin{equation}
\left\{
\begin{array}{l}
\mbox{\boldmath $\sigma$ }=\mathbf{C:}\left(
\mbox{\boldmath $\varepsilon
-\varepsilon$ }^{p}\right) \\
tr\left( \mbox{\boldmath $\dot{\varepsilon}$}^{p}\right) =0\text{ \
\ \ \ \ \ \ and \ \ \ \ \ \ \ \ \ }\mbox{\boldmath
$\dot{\varepsilon}$}^{p}=\beta \frac{dev\left( \sigma \right) }{2G}\
\dot{
\vartheta}%
\end{array}%
\right.  \label{EndoFormul2}
\end{equation}%
where the trace of the plastic strain flow $\dot{%
\mbox{\boldmath$
\varepsilon$}}^{p}$ is zero, consistently with the assumption of plastic
incompressibility. $\mathbf{C}=\left( K-2G/3\right) \mathbf{1\otimes 1}+2G%
\mathbf{I}$ is the elasticity fourth-order tensor for isotropic materials; $%
\mathbf{1}$ is the second-order identity tensor; $\mathbf{I}$ is the
fourth-order identity tensor and $\mathbf{\otimes }$ represents the tensor
product.

\subsection{Flow rules of plastically incompressible D-EC and DD-EC models}

\begin{figure}[tbp]
\begin{center}
\includegraphics[width=8.5cm]{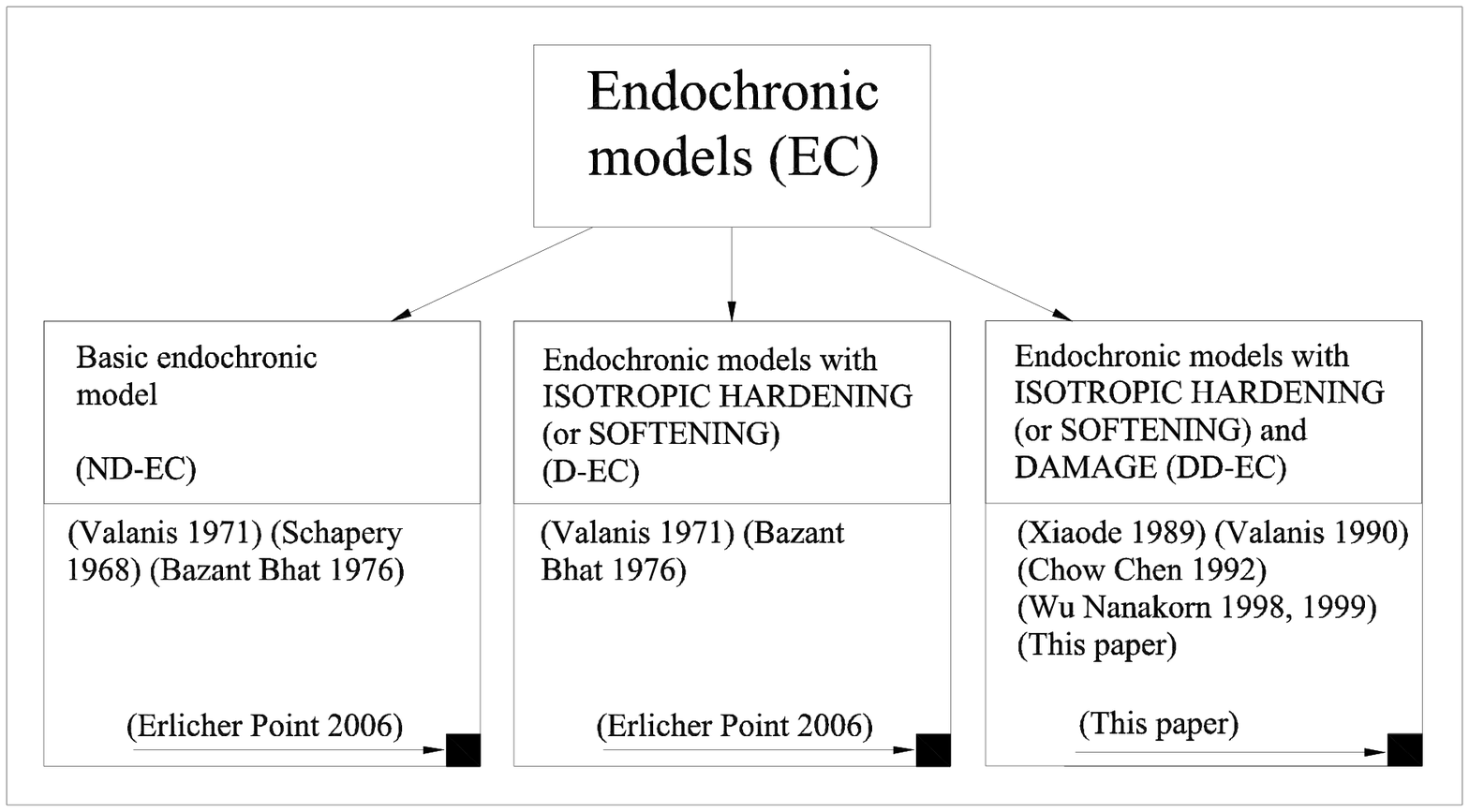}
\end{center}
\caption{Classification of Endochronic models and most relevant references.
The black square indicates that a thermodynamic formulation based on a
suited pseudo-potential was found for the associated group of models.}
\label{scheme_Endo}
\end{figure}

An endochronic model with isotropic hardening or softening with plastically
incompressible flow is defined as follows:
\begin{equation}
\left\{
\begin{array}{l}
tr\left( \mathbf{\mbox{\boldmath $\sigma$ }}\right) =3K\text{ }tr\left(
\mbox{\boldmath
$\varepsilon$ }\right) \text{ , \ \ \ \ \ \ \ \ \ }dev\left(
\mbox{\boldmath
$\sigma$ }\right) =\mathbf{z} \\
\mathbf{\dot{z}}=2G\text{ }dev\left( \mbox{\boldmath
$\dot{\varepsilon}$}\right) -\beta \text{ }\mathbf{z}\text{ }\dot{\vartheta}%
\text{ \ \ \ \ }\ \ \ \ \ \text{with }\dot{\vartheta}=\frac{\dot{\zeta}}{g}%
\end{array}%
\right.   \label{EndoFormul1}
\end{equation}%
where $g>0$ is called the \emph{hardening-softening function} %
\citep{Bazant78}. As stated by its name, the function $g$ introduces
isotropic hardening (or softening), which distinguishes this model (D-EC)
from the basic ones presented in the previous section and indicated as ND-EC
(see Figure \ref{scheme_Endo}). In the classical endochronic formulations, $g
$ is a function of $\zeta $, where $\zeta $ is the \emph{intrinsic time
measure}. A standard choice is $\dot{\zeta}=\left\Vert dev\left(
\mbox{\boldmath
$\dot{\varepsilon}$}\right) \right\Vert $ according to \citet{Valanis71}.
Another more general definition, leading to a cyclic behavior similar to
that of the Prandtl-Reuss model \citep{Lemaitre90engl} when the positive
parameter $n$ is large enough, reads
\begin{equation}
\dot{\zeta}=\left( 1+\frac{\gamma }{\beta }sgn\left(
\mathbf{z}:dev\left(\mbox{\boldmath $\dot{\varepsilon}$}\right)
\right) \right) \left\vert \mathbf{z}:dev\left( \mbox{\boldmath
$\dot{\varepsilon}$}\right) \right\vert \left\Vert
\mathbf{z}\right\Vert ^{n-2}  \label{zetan}
\end{equation}
with $\gamma \in \left[ -\beta ,\beta \right] $ in order to ensure
the non-negativity of $\dot{\zeta}$; $sgn$ is the signum function.
An important difference between (\ref{zetan}) and the standard
definition $\dot{\zeta}=\left\Vert dev\left( \mbox{\boldmath
$\dot{\varepsilon}$}\right) \right\Vert $ is related to the product
$\mathbf{z}:dev\left( \mbox{\boldmath $\dot{\varepsilon}$}\right)$,
entailing $\dot{\zeta}=0$ when the deviatoric strain increment is
orthogonal to the stress. However, $\dot{\zeta}$ can be different
from zero during unloading, i.e. when $\mathbf{z}:dev\left(
\mbox{\boldmath $\dot{\varepsilon}$}\right)<0$. Eq. (\ref {zetan})
shows that $\gamma $ affects the difference between the loading and
unloading values of the intrinsic time increment at a given stress
$\mathbf{z}$. In particular, when $\gamma =\beta $ these increments
are zero during unloading, while $\gamma $ close to (and greater
than) $-\beta $ leads to relatively small increments during loading,
while $\dot{\zeta}$ is relatively large during unloading. The
influence of $n$ on the endochronic model behavior is discussed in
the last Section, with reference to the strain accumulation and the
stress relaxation effects. According to (\ref{EndoFormul1}) and
(\ref{zetan}) and assuming $\beta +\gamma
>0$, the norm of the tensor $\mathbf{z}\left( t\right) $ is bounded
as follows:
\begin{equation}
\left\Vert \mathbf{z}\left( t\right) \right\Vert =\left\Vert
dev\left(\mbox{\boldmath $\sigma$}\left(t\right)\right) \right\Vert
<\sigma _{u}=\left( \frac{2G}{\beta +\gamma }\right) ^{\frac{1}{n}}
\label{sigmau}
\end{equation}
for $t>0$, provided that $\left\Vert \mathbf{z}\left( 0\right)
\right\Vert <\sigma _{u}$. This inequality proves that a limit
strength value exists and only concerns the deviatoric part of the
stress $\mbox{\boldmath $\sigma$}$, consistently with the plastic
incompressibility requirement. Eq. (\ref{sigmau}) also shows that
this bounding stress depends on the parameters $ \beta ,$ $\gamma$
and $n$.

The expression (\ref{EndoFormul1}) is equivalent to
\begin{equation}
\left\{
\begin{array}{l}
\mbox{\boldmath $\sigma$ }=\mathbf{C:}\left(
\mbox{\boldmath $\varepsilon
-\varepsilon$ }^{p}\right)  \\
tr\left( \mbox{\boldmath $\dot{\varepsilon}$}^{p}\right) =0\text{ \
\ \ \ \ \ \ and \ \ \ \ \ \ \ \ \ }\mbox{\boldmath
$\dot{\varepsilon}$}^{p}=\frac{dev\left( \mathbf{\sigma }\right)
}{2G/\beta} \ \frac{\dot{\zeta}}{g}
\end{array}
\right.   \label{EndoFormulGen1}
\end{equation}
From the last relationship in (\ref{EndoFormulGen1}), it appears
that the parameters $\beta ,\gamma $\ and $n$ introduced in
(\ref{zetan}) affect the amplitude of the plastic strain flow, while
the direction is always that of $dev\left( \mbox{\boldmath
$\sigma$}\right) $.

A larger class of endochronic models can be defined by the following
relationships
\begin{equation}
\left\{
\begin{array}{l}
\mbox{\boldmath $\sigma$ }=\left( 1-D\right) \mathbf{C:}\left(
\mbox{\boldmath
$\varepsilon -\varepsilon$ }^{p}\right) \\
tr\left( \mbox{\boldmath $\dot{\varepsilon}$}^{p}\right) =0\text{ \ \ \ \ \
\ \ and \ \ \ \ \ \ \ \ \ }\mbox{\boldmath
$\dot{\varepsilon}$}^{p}=\frac{1}{1-D}\frac{dev\left( \mathbf{\sigma }%
\right) }{2G/\beta }\ \frac{\dot{\zeta}}{g}%
\end{array}%
\right.  \label{endoGen1}
\end{equation}%
where $D$ is a scalar variable introducing isotropic damage. The plasticity
model with damage defined by (\ref{endoGen1}) is named here the \emph{%
extended endochronic model} and it belongs to the class of DD-EC models, as
indicated in Figure \ref{scheme_Endo}. Note that the stress is defined by
introducing the factor $\left( 1-D\right) $, consistently with the
definition of \emph{effective stress} and the \emph{principle of strain
equivalence} \citep{Lemaitre90engl}. Moreover, it can be observed that the
relationships (\ref{endoGen1}) are equivalent to
\begin{equation}
\left\{
\begin{array}{l}
tr\left( \mathbf{\mbox{\boldmath $\sigma$}}\right) =\left( 1-D\right) 3K%
\text{ }tr\left( \mbox{\boldmath $\varepsilon$}\right) \text{ , \ \ \ \ \ \
\ }dev\left( \mbox{\boldmath
$\sigma$}\right) =\mathbf{z} \\
\mathbf{\dot{z}}=\left( 1-D\right) \text{ }2G\text{ }dev\left( %
\mbox{\boldmath $\dot{\varepsilon}$}\right) -\beta \text{ }\mathbf{z}\text{ }%
\dot{\vartheta}-\dot{D}\text{ }\frac{\mathbf{z}}{1-D}\text{ \ \ \ \ }\ \ \ \
\ \text{with }\dot{\vartheta}=\frac{\dot{\zeta}}{g}%
\end{array}%
\right.  \label{sigDam}
\end{equation}%
which can be compared with (\ref{EndoFormul1}).

A possible choice for $\dot{\zeta}$ is given by
\begin{equation}
\dot{\zeta}=\left( 1+\frac{\gamma }{\beta }sgn\left( \mathbf{z}:dev\left(%
\mbox{\boldmath $\dot{\varepsilon}$}\right) \right) \right) \left\vert
\mathbf{z}:dev\left( \mbox{\boldmath
$\dot{\varepsilon}$}\right) \right\vert \left\Vert \mathbf{z}\right\Vert
^{n-2}\left( 1-D\right) ^{1-n}  \label{zetanDam}
\end{equation}
which represents a direct generalization of (\ref{zetan}): the last factor
depending on $D$ and $n$ is introduced in order to have an intrinsic time
depending on the effective stress instead of the actual one, consistently
with the strain equivalence postulate. An elastic with damage model can be
defined by assuming $\dot{\zeta}=0$. In the authors' knowledge, the notions
of pseudo-potential and loading surface were never applied to the extended
endochronic theory; therefore, these aspects are analyzed in detail in the
next section.

\section{A thermodynamic framework for the extended endochronic theory}

\label{thermoEndo}

The aim of this section is to define the Helmholtz free energy and the
pseudo-potential leading to the flow rules (\ref{endoGen1}) or,
equivalently, (\ref{sigDam}). Under the assumption of isothermal and small
transformations, the Helmholtz free energy density is chosen as follows:
\begin{equation}
\Psi =\Psi \left( \mathbf{v}\right) =\psi \left(
\mbox{\boldmath
$\varepsilon$},\mbox{\boldmath $\varepsilon$}^{p}\mathbf{,}\zeta \mathbf{,}
D\right) +\mathbb{I}_{\mathbb{H}}\left( \mbox{\boldmath
$\varepsilon$},\mbox{\boldmath $\varepsilon$}^{p}\mathbf{,}\zeta \mathbf{,}
D\right)  \label{pot0}
\end{equation}
where $\mathbf{v}=\left( \mbox{\boldmath
$\varepsilon$},\mbox{\boldmath $\varepsilon$}^{p}\mathbf{,}\zeta \mathbf{,}
D\right) $ is the vector of state variables; $\mbox{\boldmath$ \varepsilon$}$%
, $\mbox{\boldmath$\varepsilon$}^{p}$ and $D$ were previously defined; $%
\zeta $ is a scalar internal variable associated with isotropic hardening.
For all the state variables, an initial zero value is assumed. The choice of
$\zeta $ to indicate an internal variable might seem misleading, since the
symbol $\zeta $ was also used in (\ref{EndoFormul1})-(\ref{sigDam}) to
define the \emph{intrinsic time measure}. However, as it will be seen
hereafter, this choice is the proper one, as for endochronic models, $\zeta $
has simultaneously both meanings; $\psi $ is the regular part of the
Helmholtz energy; $\mathbb{I}_{\mathbb{H}}$ is the indicator function of the
closed set $\mathbb{H}$: by definition, an indicator function is equal to $0$
inside $\mathbb{H}$ and equal to $+\infty $ outside \citep{Rockafellar69};
the set $\mathbb{H}$ indicates the admissibility domain for the state
variables $\mathbf{v}$ and should be introduced every time some conditions
on state variables are to be imposed: for instance, it is equal to the
interval $D\in \lbrack 0,1]$ in order to impose the admissible values for
the damage variable \citep{Fremond2002}.

Once $\Psi $ is known, the \emph{non-dissipative} thermodynamic forces $%
\mathbf{q}^{nd}=(\mbox{\boldmath
$\sigma$}^{nd},\mbox{\boldmath$\tau$}^{nd},R^{nd},Y^{nd})$ are defined as
the gradient of $\psi \left( \mbox{\boldmath
$\varepsilon$},\mbox{\boldmath $\varepsilon$}^{p}\mathbf{,}\zeta \mathbf{,}
D\right) $:
\begin{equation}
\mbox{\boldmath $\sigma$}^{nd}:=\frac{\partial \psi }{\partial %
\mbox{\boldmath $\varepsilon$}},\text{ \ \ }\mbox{\boldmath$\tau$}^{nd}:=
\frac{\partial \psi }{\partial \mbox{\boldmath $\varepsilon$}^{p}},\text{ \
\ }R^{nd}:=\frac{\partial \psi }{\partial \zeta },\text{ \ \ }Y^{nd}:=\frac{
\partial \psi }{\partial D}
\end{equation}
while the \emph{non-dissipative reaction forces} $\mathbf{q}^{ndr}=( %
\mbox{\boldmath$\sigma$}^{ndr},\mbox{\boldmath$\tau$}^{ndr},R^{ndr},Y^{ndr})$
are given by
\begin{equation}
\left( \mbox{\boldmath$\sigma$}^{ndr},\mbox{\boldmath$\tau$}
^{ndr},R^{ndr},Y^{ndr}\right) \in \partial \mathbb{I}_{\mathbb{H}}\left(
\mbox{\boldmath
$\varepsilon$},\mbox{\boldmath $\varepsilon$}^{p}\mathbf{,}\zeta \mathbf{,}
D\right)
\end{equation}
where $\partial $ is the \emph{sub-differential} operator %
\citep{Rockafellar69}. If the constraints imposed by $\mathbb{H}$ are
fulfilled, the indicator function $\mathbb{I}_{\mathbb{H}}$ is zero and $%
\Psi \left( t\right) =\psi \left( t\right)$. This entails the identity of
the time-derivatives, viz. $\dot{\Psi}\left( t\right) =\dot{\psi}\left(
t\right) +\mathbf{q}^{ndr}\cdot \mathbf{\dot{v}}=\dot{\psi}\left( t\right)$.
In other words, one has $\mathbf{q}^{ndr}\cdot \mathbf{\dot{v}=}0$ for every
instant $t$ \citep{Fremond2002}.

Due to the assumptions of isothermal and small transformations, the
expression of the second principle reads:
\begin{equation}
\Phi _{1}\left( t\right) =\mbox{\boldmath $\sigma
:\dot{\varepsilon}$}-\dot{\psi}\geq 0  \label{2ndprinc}
\end{equation}%
(\ref{2ndprinc}) states that the \emph{intrinsic }(or \emph{mechanical})
\emph{dissipation} $\Phi _{1}$ has to be non-negative. Introducing the \emph{%
dissipative} thermodynamic forces $\mathbf{q}^{d}=(\mbox{\boldmath
$\sigma$}^{d},\mbox{\boldmath$\tau$}^{d},R^{d},Y^{d})$ as
\begin{equation}
\mbox{\boldmath $\sigma$}^{d}:=\mbox{\boldmath $\sigma
-\sigma$}^{nd}-\mbox{\boldmath$\sigma$}^{ndr}\text{ ,\ \ \ \ \ } %
\mbox{\boldmath$\tau$}^{d}:=-\mbox{\boldmath$\tau$}^{nd}-%
\mbox{\boldmath$
\tau$}^{ndr},\text{\ \ \ \ }R^{d}:=-R^{nd}-R^{ndr},\text{\ \ \ \ }
Y^{d}:=-Y^{nd}-Y^{ndr}  \label{StateThermoForce}
\end{equation}
and substituting (\ref{StateThermoForce}) in (\ref{2ndprinc}), one obtains:
\begin{equation}
\Phi _{1}\left( t\right) =\mbox{\boldmath $\sigma$}^{d}:
\mbox{\boldmath
$\dot{\varepsilon}$}+\mbox{\boldmath$\tau$}^{d}:
\mbox{\boldmath
$\dot{\varepsilon}$}^{p}+R^{d}\dot{\zeta}+Y^{d}\text{ }\dot{D}\text{ }
\mathbf{\geq }\text{ }0  \label{2ndprinc1}
\end{equation}

In order to fulfill (\ref{2ndprinc1}), the flows of the state variables $%
\mbox{\boldmath $\dot{\varepsilon}$},\mbox{\boldmath
$\dot{\varepsilon}$}^{p}\mathbf{,}\dot{\zeta}$ and $\dot{D}$ have to be
suitably correlated with the \emph{dissipative} thermodynamic forces $%
\mbox{\boldmath
$\sigma$}^{d},\mbox{\boldmath$\tau$}^{d},R^{d}$ and $Y^{d}$. Therefore, some
additional \emph{complementarity rules} need to be defined: usually, a
scalar non-negative function called \emph{pseudo-potential}
\begin{equation}
\phi =\phi \left( \mathbf{\dot{v}}^{\prime };\mathbf{v};
\mbox{\boldmath$
\rho$}\right) \text{ \ \ such that \ \ \ \ \ \ }\phi \left( \mathbf{0};
\mathbf{v};\mbox{\boldmath$\rho$}\right) =0\text{ \ \ \ for all }\mathbf{v}
\text{ and }\mbox{\boldmath$\rho$}
\end{equation}%
is introduced and the dissipative forces $\mathbf{q}^{d}=(%
\mbox{\boldmath$\sigma$}^{d},\mbox{\boldmath$\tau$}^{d},R^{d},Y^{d})$ are
derived imposing the so-called \emph{generalized normality assumption} on
it. Equivalently, one can define the flow rules $\mathbf{\dot{v}}$ by
imposing the generalized normality assumption on the \emph{dual}
pseudo-potential $\phi ^{\ast }$, which is the Legendre-Fenchel transform of
$\phi $ \citep{Rockafellar69}. This last method will be explicitly exploited
herein. The generic flow $\mathbf{\dot{v}}^{\prime }$ is noted with "prime",
while the actual flow at the present state is noted with $\mathbf{\dot{v}}$.
As a matter of fact, the pseudo-potential is assumed to vary with the
present value of state variables $\mathbf{v}$ and with some additional
parameters collected in the vector $\mbox{\boldmath$\rho$}$. These
parameters may be any quantity related to the past history of the material %
\citep{Fremond2002}. For instance, one may have $\mbox{\boldmath$ \rho$}(
\mathbf{x})=\left( e(\mathbf{x}),\left\Vert \mbox{\boldmath$
\varepsilon$}(\mathbf{x})\right\Vert _{\max }\right) =\left( e(\mathbf{x}
),\max_{0\leq t^{\prime }\leq t}\left\Vert \mbox{\boldmath $\varepsilon$}
\left( \mathbf{x},t^{\prime }\right) \right\Vert \right) $, where $e$ is the
dissipated energy per unit volume at the point $\mathbf{x}$ of the body
volume and $\left\Vert \mbox{\boldmath $\varepsilon$}\right\Vert_{\max }$ is
the maximum (from $t^{\prime}=0$ to the present state $t^{\prime}=t$) of the
strain norm at the same point. Observe that the parameters collected in $%
\mbox{\boldmath$\rho$}$ could also be \emph{non-local}, like $\rho \left(
\mathbf{x}\right) \mathbf{=}\int_{V_{0}\left( \mathbf{x}\right) }edV$ , i.e.
the energy dissipated in a given volume $V_{0}\left( \mathbf{x}\right) $
around the point $\mathbf{x}$\ of body volume.

\begin{figure}[tbp]
\begin{center}
\includegraphics[width=13cm]{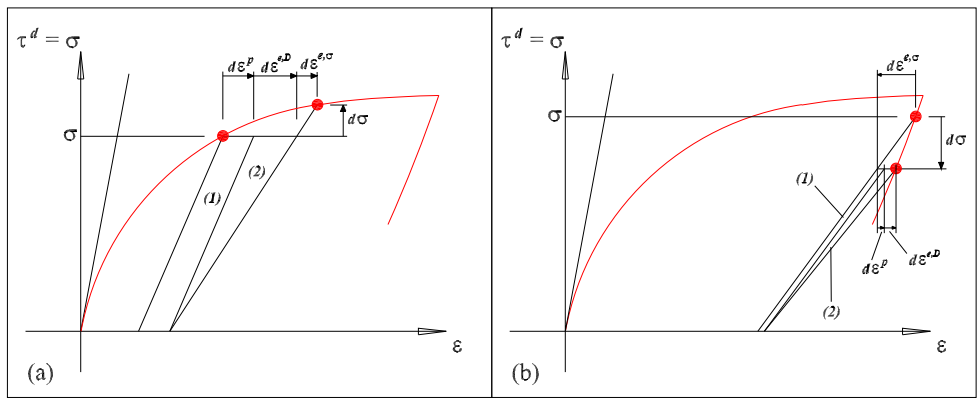}
\end{center}
\caption{The increment $de_{p}$ of the energy dissipated by plastic strain
(1) and the increment $de_{D}$ of the energy dissipated by damage (2). The
increments of the elastic and plastic strain as well as of the stress are
also schematically illustrated for a) the loading phase; and b) the
unloading phase.}
\label{Fig_dissEn}
\end{figure}

When no viscous effect occurs, the case of plasticity with damage is
recovered. This corresponds to choose a pseudo-potential $\phi $ independent
from $\mbox{\boldmath $\dot{\varepsilon}$}$, entailing $%
\mbox{\boldmath
$\sigma$}^{d}=\mathbf{0}$; for a detailed derivation of these relationships,
see, for instance, \citet{ErlicherPoint2006}. Moreover, "plastic flow may
occur without damage and damage may occur without appreciable macroscopic
plastic flow" \citep{Lemaitre90engl}. Therefore, (\ref{2ndprinc1}) with $%
\mbox{\boldmath $\sigma$}^{d}=\mathbf{0}$ "must be split in two independent
inequalities":
\begin{equation}
\dot{e}_{p}:=\mbox{\boldmath$\tau$}^{d}:\mbox{\boldmath
$\dot{\varepsilon}$}^{p}+R^{d}\text{ }\dot{\zeta}\geq 0,\text{ \ \ \ \ \ \ \
\ \ \ \ \ \ \ }\dot{e}_{D}:=Y^{d}\text{ }\dot{D}\geq 0  \label{doubleIneq}
\end{equation}
The two scalar quantities $\dot{e}_{p}$ and $\dot{e}_{D}$ respectively
define the rate of energy per unit volume dissipated by plasticity-related
phenomena and by damage phenomena; see Figure \ref{Fig_dissEn}. Their sum $%
\dot{e}=\dot{e}_{p}+\dot{e}_{D}$ is the rate of the total dissipated energy
per unit volume and coincides with the intrinsic dissipation $\Phi_{1}$. The
restrictions imposed by these two inequalities are more severe than the
original unique inequality of Clausius-Duhem (\ref{2ndprinc1}). However,
they are usually adopted as basic thermodynamic criterion for the
formulation of plasticity models with damage \citep{Lemaitre90engl}. This
assumption will be adopted hereafter. Taking into account (\ref{doubleIneq}%
), the pseudo-potential is supposed to split into two pseudo-potentials $%
\phi _{D}$ and $\phi _{p}$, respectively related to damage and plastic flow:
\begin{equation}
\phi \left( \mbox{\boldmath$\dot{\varepsilon}$}^{p^{\prime }},\dot{\zeta}
^{\prime },\dot{D}^{\prime };\mathbf{v};\mbox{\boldmath$\rho$}\right) =\phi
_{D}\left( \dot{D}^{\prime }; \mathbf{v};\mbox{\boldmath$\rho$}\right)+\phi
_{p}\left( \mbox{\boldmath$\dot{\varepsilon}$}^{p^{\prime }},\dot{\zeta}%
^{\prime };\mathbf{v};\mbox{\boldmath$\rho$}\right)
\end{equation}
In the following sections, the Helmholtz free energy, the pseudo-potentials $%
\phi _{D}$ and $\phi_{p}$, as well as their Legendre-Fenchel transforms %
\citep{Rockafellar69}, are formulated for the endochronic model with damage (%
\ref{endoGen1}).

\subsection{The Helmholtz free energy}

According to (\ref{pot0}), for the DD-EC models one has the following
Helmholtz free energy:
\begin{equation}
\Psi \left( \mathbf{v}\right) =\psi \left( \mathbf{v}\right) +\mathbb{I}_{%
\mathbb{H}}=\left( 1-D\right) \frac{1}{2}\left(
\mbox{\boldmath $\varepsilon
-\varepsilon$}^{p}\right) \mathbf{:C:}\left(
\mbox{\boldmath $\varepsilon
-\varepsilon$}^{p}\right) +\mathbb{I}_{\mathbb{H}}  \label{Helm}
\end{equation}
In this paper, two cases are considered:
\begin{equation}
\mathbb{H}=\left\{
\begin{array}{l}
\left( \mbox{\boldmath $\varepsilon,\varepsilon$}^{p},\zeta ,D\right) \ \
\text{such\ that \ \ }D\geq 0,\ D\leq 1 \\
\text{and \ \ }\left( 1-D\right) ^{s}R\left(\tilde{\mathbf{v}}, %
\mbox{\boldmath $\rho$}\right) -r_{0}\leq 0%
\end{array}
\right\}  \label{H1}
\end{equation}
where $s$ and $r_{0}$ are positive parameters; $\tilde{\mathbf{v}}=\left(%
\mbox{\boldmath $\varepsilon,\varepsilon$} ^{p},\zeta\right)$ collects all
state variables except $D$ and $R=R\left(\tilde{\mathbf{v}},%
\mbox{\boldmath
$\rho$}\right)$ is a non-negative function called \emph{source of damage}.
The first two conditions on $D$ impose the minimum and the maximum values
for this variable. As it will be seen, the third condition in (\ref{H1}) is
strictly related to the definition of the damage limit surface. The second
case is characterized by a different assumption:
\begin{equation}
\mathbb{H}=\left\{ \left( \mbox{\boldmath $\varepsilon,\varepsilon$}
^{p},\zeta ,D\right) \ \ \text{such\ that \ }D\geq 0,\ D\leq 1\right\}
\label{H2}
\end{equation}
where only the two basic inequalities on $D$ are retained.

Making use of (\ref{StateThermoForce}), (\ref{Helm}), (\ref{H1}) and of the
pseudo-potential (\ref{phi D}), i.e. the Definition 1 of $\phi _{D}$ given
in the following section, it is possible to prove that the assumption $%
\mathbf{q}^{ndr}=\mathbf{0}$ is admissible. The same holds for the model
defined by (\ref{Helm}),(\ref{H2}) and (\ref{pseudoDamMod}) (Definition 2 of
$\phi_D$). For brevity, the details of this proof, are omitted. As a result,
the non-dissipative thermodynamic forces fulfill the following
relationships:
\begin{equation}
\begin{array}{l}
\mbox{\boldmath $\sigma$}^{nd}=\frac{\partial \psi }{\partial %
\mbox{\boldmath $\varepsilon$}}=\left( 1-D\right) \text{ }\mathbf{C:}\left(
\mbox{\boldmath $\varepsilon
-\varepsilon$}^{p}\right) =\mbox{\boldmath $\sigma$}-
\mbox{\boldmath
$\sigma$}^{d}=\mbox{\boldmath $\sigma$} \\
\mbox{\boldmath$\tau$}^{nd}=\frac{\partial \psi }{\partial
\mbox{\boldmath
$\varepsilon$}^{p}}=\mathbf{-}\left( 1-D\right) \text{ }\mathbf{C:}\left(
\mbox{\boldmath
$\varepsilon -\varepsilon$}^{p}\right) =-\mbox{\boldmath$\tau$}^{d} \\
R^{nd}=\frac{\partial \psi }{\partial \zeta }=0=-R^{d} \\
Y^{nd}=\frac{\partial \psi }{\partial D}=-\frac{1}{2}\left(
\mbox{\boldmath
$\varepsilon-\varepsilon$}^{p}\right) \mathbf{:C:}\left(
\mbox{\boldmath
$\varepsilon-\varepsilon$}^{p}\right) =-Y^{d}%
\end{array}
\label{nondissThermForces}
\end{equation}
Moreover using (\ref{nondissThermForces}) and supposing $D<1$, the energy
dissipation rate reads
\begin{equation}
\begin{array}{l}
\dot{e}=\dot{e}_{p}+\dot{e}_{D}=\mbox{\boldmath $\sigma$}:
\mbox{\boldmath
$\dot{\varepsilon}$}^{p}+\frac{1}{2}\mbox{\boldmath
$\sigma$}:\frac{\mathbf{C}^{-1}:\mbox{\boldmath $\sigma$}}{\left( 1-D\right)
^{2}}\dot{D} \\
=dev\left( \mbox{\boldmath $\sigma$}\right) :dev\left(
\mbox{\boldmath
$\dot{\varepsilon}$}^{p}\right) +\frac{tr\left( \sigma \right) }{3}tr\left(
\mbox{\boldmath
$\dot{\varepsilon}$}^{p}\right) +\frac{1}{2}\left( \frac{dev\left( \sigma
\right) \mathbf{:}dev\left( \sigma \right) }{2G}\mathbf{+}\frac{\left(
tr\left( \mathbf{\sigma }\right) \right) ^{2}}{9K}\right) \text{ }\frac{\dot{
D}}{\left( 1-D\right) ^{2}}%
\end{array}
\label{dissEner}
\end{equation}
Taking the time-derivative of (\ref{nondissThermForces})$_{1}$ and recalling
that no viscous effect is considered ($\mbox{\boldmath$\sigma$}^{d}=\mathbf{0%
}$), one obtains
\begin{equation}
\mbox{\boldmath$\dot{\varepsilon}$}=\mbox{\boldmath
$\dot{\varepsilon}$}^{p}+\mbox{\boldmath
$\dot{\varepsilon}$}^{e}=\mbox{\boldmath
$\dot{\varepsilon}$}^{p}+\mathbf{C}^{-1}:\frac{
\mbox{\boldmath
$\dot{\sigma}$}}{1-D}+\mathbf{C}^{-1}:\frac{\mbox{\boldmath
$\sigma$}\dot{D}}{\left( 1-D\right) ^{2}}:=
\mbox{\boldmath
$\dot{\varepsilon}$}^{p}+\mbox{\boldmath
$\dot{\varepsilon}$}^{e,\sigma }+\mbox{\boldmath
$\dot{\varepsilon}$}^{e,D}
\end{equation}
where $\mbox{\boldmath $\dot{\varepsilon}$}^{e,\sigma }$ is an elastic
strain flow at constant damage and $\mbox{\boldmath
$\dot{\varepsilon}$}^{e,D}$ is an elastic strain flow at constant stress. It
follows that $\dot{e}_{D}=1/2\left(\mbox{\boldmath
$\sigma$}:\mbox{\boldmath $\dot{\varepsilon}$}^{e,D}\right)$. Note that in
general the endochronic theory may present non-zero energy rates $\dot{e}%
_{p} $ and $\dot{e}_{D}$ also during unloading phases; see Figure \ref%
{Fig_dissEn}b in this respect.

\subsection{The pseudo-potential for the damage flow}

The formalism of the loading function $f_{D}$, as well as the
pseudo-potential $\phi_{D}^{\ast }=\mathbb{I}_{f_{D}\leq 0}$ , can be used
to express the damage evolution
\citep{Lemaitre90engl,
Salari2004, Nedjar2001, Fremond2002}. We present herein a well-known damage
evolution rule by using both pseudo-potentials $\phi_{D}$ and its dual $%
\phi_{D}^{\ast}$. Then, a discussion is done about a novel pseudo-potential
leading to a damage evolution where $\dot{D}$ may be different from 0 also
during unloading phases. In detail, the main difference between the two
cases is related to the role of the damage limit surface. Standard damage
evolution rules, viz. Definition 1, are characterized by the possibility for
the actual state point to be inside the damage domain delimited by this
limit surface; in this situation and in particular during unloading phases,
damage increments are null. Conversely, in the damage evolution which we
propose here, i.e. Definition 2, the present state point is forced to be
always on the damage limit surface also during unloading phases.

\subsubsection{Definition 1 of $\protect\phi _{D}$}

Let us begin with the following pseudo-potential, associated with the
Helmholtz free energy (\ref{Helm})-(\ref{H1}):
\begin{equation}
\phi _{D}\left( \dot{D}^{\prime };\mathbf{v};\mbox{\boldmath$\rho$}\right) = %
\left[ \frac{1}{2}\left( \mbox{\boldmath $\varepsilon -\varepsilon$}
^{p}\right) :\mathbf{C:}\left( \mbox{\boldmath
$\varepsilon-\varepsilon$} ^{p}\right) -\left( 1-D\right) ^{s}R\left( \tilde{
\mathbf{v}}, \mbox{\boldmath$\rho$}\right) +r_{0}\right] \text{ }\dot{D}%
^{\prime }+ \mathbb{I}_{\mathbb{D}_{D}}\left( \dot{D}^{\prime }\right)
\label{phi D}
\end{equation}
and
\begin{equation}
\mathbb{D}_{D}=\left\{ \dot{D}^{\prime }\text{ such that }\dot{D}^{\prime
}\geq 0\right\}  \label{effectiveDd}
\end{equation}
The pseudo-potential $\phi _{D}$ is the sum of a \emph{regular} part,
proportional to $\dot{D}^{\prime }$ and of the indicator function $\mathbb{I}
_{\mathbb{D}_{D}}$. The term multiplying $\dot{D}^{\prime }$ in the regular
part of $\phi _{D}$ is always non-negative, by virtue of the third condition
defining $\mathbb{H}$ in (\ref{H1}). The regular part of $\phi _{D}$,
considered for the actual flow $\dot{D}^{\prime }=\dot{D}$, represents the
rate of dissipated energy $\dot{e}_{D}$. (\ref{phi D})-(\ref{effectiveDd})
allow a large number of standard damage evolution rules to be represented,
according to the specific definition of $R\left(\tilde{\mathbf{v}},%
\mbox{\boldmath$\rho$}\right) $. An interesting example is
\begin{equation}
R=R\left( \mbox{\boldmath$\varepsilon$}\right) =\frac{1}{2}\left( 2G %
\mbox{\boldmath $\varepsilon$}^{e,+}:\mbox{\boldmath
$\varepsilon$}^{e,+}+\lambda \left( \left[ tr\left(
\mbox{\boldmath
$\varepsilon$}^e\right) \right]^{+}\right) ^{2}\right)  \label{FreDam}
\end{equation}
where $\lambda =K-2G/3$ is the Lam\'{e} constant. For a scalar $x$, $\left[x%
\right] ^{+}:=\left\langle x\right\rangle $, where $\left\langle
{}\right\rangle $ are the McCauley brackets. The positive part $%
\mbox{\boldmath$\varepsilon$}^{e,+}$ of the tensor $%
\mbox{\boldmath
$\varepsilon$}^{e}$ is obtained after diagonalisation. Other definitions for
$R$ can be adopted; see, among others, \citet{Nedjar2001} and %
\citet{Salari2004}.

In order to derive the damage flow, it is convenient to consider the
Legendre-Fenchel transform of $\phi_{D}$, which reads:
\begin{equation}
\begin{array}{l}
\phi _{D}^{\ast }\left( Y^{d^{\prime
}};\mathbf{v};\mbox{\boldmath$\rho$} \right) =\sup_{\dot{D}^{\prime
}\in \mathbb{D}_{D}}\left[ \left( Y^{d^{\prime }}-\frac{1}{2}\left(
\mbox{\boldmath $\varepsilon -\varepsilon$} ^{p}\right)
:\mathbf{C:}\left( \mbox{\boldmath $\varepsilon
-\varepsilon$}^{p}\right) +\left( 1-D\right) ^{s}R\left(
\tilde{\mathbf{v}},
\mbox{\boldmath$\rho$}\right) -r_{0}\right) \dot{D}^{\prime }\right] \\
=\mathbb{I}_{\mathbb{E}_{D}}\left( Y^{d^{\prime }};\mathbf{v};
\mbox{\boldmath$\rho$}\right)
\end{array}
\label{phi D star}
\end{equation}
where $\mathbb{E}_{D}\mathbb{=}\left\{ Y^{d^{\prime }}\text{ such
that \ \ } f_{D}\left( Y^{d^{\prime
}};\mathbf{v};\mbox{\boldmath$\rho$}\right) \leq 0\right\} $ is the
damage loading domain and the corresponding loading function is:
\begin{equation}
f_{D}\left( Y^{d^{\prime }};\mathbf{v};\mbox{\boldmath$\rho$}\right)
=Y^{d^{\prime }}-\frac{1}{2}\left( \mbox{\boldmath $\varepsilon
-\varepsilon$}^{p}\right) :\mathbf{\ C:}\left( \mbox{\boldmath
$\varepsilon -\varepsilon$}^{p}\right) +\left( 1-D\right)
^{s}R\left( \tilde{\mathbf{v}},\mbox{\boldmath$\rho$}\right)
-r_{0}:=Y^{d^{\prime }}-Y_{\max }^{d^{\prime }}  \label{fD}
\end{equation}
By using the normality assumption, the relevant damage flow rule reads
\begin{equation}
\begin{array}{l}
\dot{D}=\dot{\lambda}_{D}\frac{\partial f_{D}\left( Y^{d}\right) }{\partial
Y^{d^{\prime }}}=\dot{\lambda}_{D} \\
\text{with }f_{D}\left(
Y^{d};\mathbf{v};\mbox{\boldmath$\rho$}\right) \dot{
\lambda}_{D}=0,\text{ \ \ }\dot{\lambda}_{D}\geq 0,\text{ \ \ \ \ }
f_{D}\left( Y^{d};\mathbf{v};\mbox{\boldmath$\rho$}\right) \leq 0.
\end{array}
\label{dam1KuhnTuck}
\end{equation}
At the actual state, it holds $Y^{d^{\prime }}=Y^{d}=-Y^{nd}$, and
therefore:
\begin{equation}
f_{D}=f_{D}\left( Y^{d};\mathbf{v};\mbox{\boldmath$\rho$}\right) =\left(
1-D\right) ^{s}R\left( \tilde{\mathbf{v}},\mbox{\boldmath$\rho$}\right)
-r_{0}\leq 0,
\end{equation}
which is the damage limit surface, but also is one of the conditions
defining the set $\mathbb{H}$. It becomes evident that the positive
constant $r_{0}$ is the initial damage threshold. The Kuhn-Tucker
conditions state that $f_{D}<0$ implies no damage increment, while
$f_{D}\left( Y^{d};\mathbf{v};\mbox{\boldmath$\rho$}\right) =0$
corresponds to a damage increment which can be computed by enforcing
the consistency condition:
\begin{equation}
\dot{f}_{D}=\left( 1-D\right) ^{s}\left( \frac{\partial R\left(
\tilde{ \mathbf{v}},\mbox{\boldmath$\rho$}\right) }{\partial
\tilde{\mathbf{v}}} \mathbf{\dot{\tilde{v}}+}\frac{\partial R\left(
\tilde{\mathbf{v}}, \mbox{\boldmath$\rho$}\right) }{\partial
\mbox{\boldmath$\rho$}} \mbox{\boldmath$\dot{\rho}$}\right) -R\left(
\tilde{\mathbf{v}}, \mbox{\boldmath$\rho$}\right) s\left( 1-D\right)
^{s-1}\dot{D}=0
\end{equation}
leading to the explicit expression of the damage flow
\begin{equation}
\dot{D}=H\left( f_{D}\right) \left( \frac{\partial R\left(
\tilde{\mathbf{v}} ,\mbox{\boldmath$\rho$}\right) }{\partial
\tilde{\mathbf{v}}}\mathbf{\dot{ \tilde{v}}+}\frac{\partial R\left(
\tilde{\mathbf{v}},\mbox{\boldmath$\rho$} \right) }{\partial
\mbox{\boldmath$\rho$}}\mbox{\boldmath$\dot{\rho}$} \right)
\frac{1-D}{s\text{ }R\left(
\tilde{\mathbf{v}},\mbox{\boldmath$\rho$} \right) },
\label{damageFlowGen}
\end{equation}
where $H$ is the Heaviside function. The presence of the Heaviside function
in the damage flow definition indicates that damage increments are zero
during unloading phases. Note that (\ref{damageFlowGen}) entails that the
limit condition $D=1$ is never reached.

\begin{figure}[tbp]
\begin{center}
\includegraphics[width=11cm]{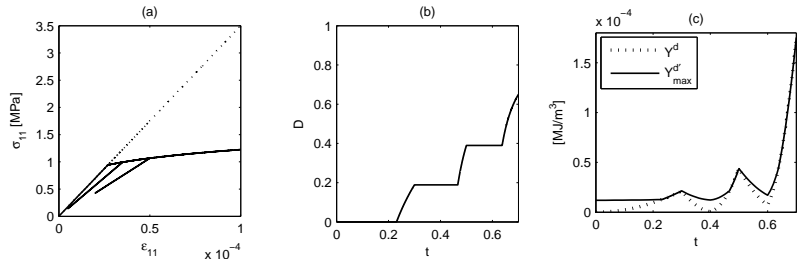}
\end{center}
\caption{Elastic with damage model. Uniaxial behavior. (a) The stress-strain
evolution. The dotted line indicates the effective stress $\protect\sigma%
_{11}/(1-D)$. (b) The damage evolution, with the damage increasing
only during loading phases. (c) Time-evolution of $Y^d$, i.e. the
actual value of the thermodynamic force $Y^{d^{\prime}}$, and of
$Y^{d^{\prime}}_{max}$, i.e. the upper bound of $Y^{d^{\prime}}$
defined in (\ref{fD}).} \label{Fig_DamFremEl11}
\end{figure}

Figure \ref{Fig_DamFremEl11} illustrates some loading-unloading
cycles of an elastic with damage model ($\dot{\zeta}=0$). The
uniaxial stress is considered, viz. all the components of the Cauchy
tensor are supposed to be null, except $\sigma _{11}$. The parameter
values represent a hypothetical material for which the Young modulus
$E=35000$ $MPa$ and the Poisson ratio $\nu =0.18$ are close to those
of concrete; damage is defined by (\ref{H1}) and (\ref{FreDam}),
with $s=2.5$ and $r_{0}=1.2e-05$ $MJ/m^{3}$; see also the numerical
examples in \cite{Nedjar2001}. Together with the stress-strain and
damage evolution of this model, Figure \ref{Fig_DamFremEl11} depicts
the evolution of $Y^{d}=1/2\left( \mbox{\boldmath $\varepsilon
-\varepsilon$} ^{p}\right) :\mathbf{C:}\left( \mbox{\boldmath
$\varepsilon -\varepsilon$ }^{p}\right) $, i.e. the actual value of
$Y^{d^{\prime }}$, and of the quantity $Y_{\max }^{d^{\prime
}}=1/2\left( \mbox{\boldmath $\varepsilon-\varepsilon$ }^{p}\right)
:\mathbf{C:}\left( \mbox{\boldmath $\varepsilon -\varepsilon$
}^{p}\right) +r_{0}-(1-D)^{s}R\left( \mbox{\boldmath
$\varepsilon$}\right) $, defining the upper limit of $Y^{d^{\prime
}}$ according to (\ref{fD}). When these two curves are superposed,
the damage increases.

\subsubsection{Definition 2 of $\protect\phi _{D}$}

Unfortunately, a definition of $\dot{D}$ of the type
(\ref{damageFlowGen}), deriving from the pseudo-potential (\ref{phi
D}) and the condition (\ref{H1}), is not able to represent the case
of damage increasing during both loading and unloading phases, owing
to the condition $f_{D}\leq 0$. We recall that the case of damage
increasing during unloading may occur in Bouc-Wen models with
stiffness degradation \citep{ErlicherBursi2007}. A damage
pseudo-potential, simpler than (\ref{phi D}), is more suited:
\begin{equation}
\phi _{D}\left( \dot{D}^{\prime };\mathbf{v}\right) =\left[
\frac{1}{2} \left( \mbox{\boldmath $\varepsilon -\varepsilon$
}^{p}\right) :\mathbf{C:}\left( \mbox{\boldmath $\varepsilon
-\varepsilon$}^{p}\right) \right] \text{ }\dot{D}^{\prime
}+\mathbb{I}_{\mathbb{D}_{D}}\left( \dot{D}^{\prime }\right)
\label{pseudoDamMod}
\end{equation}%
with $\mathbb{D}_{D}$ still provided by (\ref{effectiveDd}) and with
the conditions on the damage state variable defined in (\ref{H2}).
As already observed, it is possible to prove that the assumption
$\mathbf{q}^{ndr}=\mathbf{0}$ is admissible also for this Definition
2 of the damage pseudo-potential. The dual pseudo-potential becomes
$\phi _{D}^{\ast }\left( Y^{d^{\prime }};\mathbf{v}\right)
=\mathbb{I}_{\mathbb{E}_{D}}\left( Y^{d^{\prime }};\mathbf{\ \ \
v}\right) $ where $\mathbb{E}_{D}\mathbb{=} \left\{ Y^{d^{\prime
}}\text{ such that \ }f_{D}\left( Y^{d^{\prime }}; \mathbf{v}\right)
\leq 0\right\} $ is the corresponding damage loading domain, with
the damage loading function
\begin{equation}
f_{D}\left( Y^{d^{\prime }};\mathbf{v}\right) =Y^{d^{\prime
}}-\frac{1}{2} \left( \mbox{\boldmath $\varepsilon -\varepsilon$
}^{p}\right) :\mathbf{C:}\left( \mbox{\boldmath $\varepsilon
-\varepsilon$ }^{p}\right):=Y^{d^{\prime}}-Y^{d^{\prime}}_{max}
\label{fD1}
\end{equation}
At the actual state, $Y^{d^{\prime }}=Y^{d}=-Y^{nd}$ and therefore
$f_{D}\left( Y^{d};\mathbf{v}\right) =0$ at every instant.
Therefore, the relationships (\ref{dam1KuhnTuck}) reduce to
$\dot{D}=\dot{\lambda} _{D}\partial f_{D}/\partial Y^{d^{\prime
}}=\dot{\lambda}_{D}$, with $\dot{\lambda}_{D}\geq 0$. Moreover,
$\dot{D}=\dot{\lambda}_{D}$ can no longer be computed by the
consistency condition, fulfilled as an identity at every instant.
Hence, it must be rather defined by an additional condition. Any
definition ensuring rate-independence, consistent with (\ref{H2})
and fulfilling $\dot{D}=\dot{\lambda}_{D}\geq 0$ is admissible, even
though is characterized by non-zero damage increments during
unloading phases.

\subsection{The pseudo-potential for the plastic flow}

The usual method to define associated plastic flows is based on the notion
of \emph{loading function}, indicated here by $f_{p}$, as well as on the
normality assumption. Another equivalent formalism is based on the use of
the dual pseudo-potential $\phi _{p}^{\ast}= \mathbb{I}_{f_{p}\leq 0}$ %
\citep{Moreau70}. A third way to formulate plasticity models is based on the
pseudo-potential $\phi _{p}$, Legendre-Fenchel conjugate of $\phi_{p}^{\ast
} $ \citep{Fremond2002, Ziegler87, Houlsby2000, ErlicherPoint2006}. The
advantage of using the formalism based on $f_{p}$ (or $\phi _{p}^{\ast }=
\mathbb{I}_{f_{p}\leq 0}$) is essentially simplicity. Moreover, when a
non-associated flow is to be defined, the simple introduction of a second
function $g_p$ called plastic potential matches this purpose. Nonetheless,
for some non-classical plasticity theories, like endochronic theory and
generalized plasticity \citep{Lubliner93}, it is not straightforward to
provide a proper definition of the loading function $f_{p}$. It was proved
by \citet{ErlicherPoint2006} that for these plasticity theories (without
damage) a way to define the loading function is to start from the definition
of the pseudo-potential $\phi_{p}$, to compute the dual potential $\phi
_{p}^{\ast }$ and then to derive $f_{p}$. An important point is the
additional dependence of $\phi_p$, and therefore of $\phi_p^{\ast}$ and the
loading function too, on the \emph{state variables}. This dependence is only
optional for standard plasticity theories but is essential both for the
endochronic theory and the generalized plasticity. Moreover, we notice that
some models with non-associated flow also admit a representation based on
the definition of a suited pseudo-potential $\phi _{p}$, depending on state
variables. The example of a non-associated Drucker-Prager model can be found
in \citet{ErlicherPoint2005}; in particular, it is shown that a suited
pseudo-potential $\phi _{p}$ leads to a \emph{modified} loading function
which plays both roles of the traditional loading function and of the
plastic potential.

\begin{figure}[tbp]
\begin{center}
\includegraphics[width=14cm]{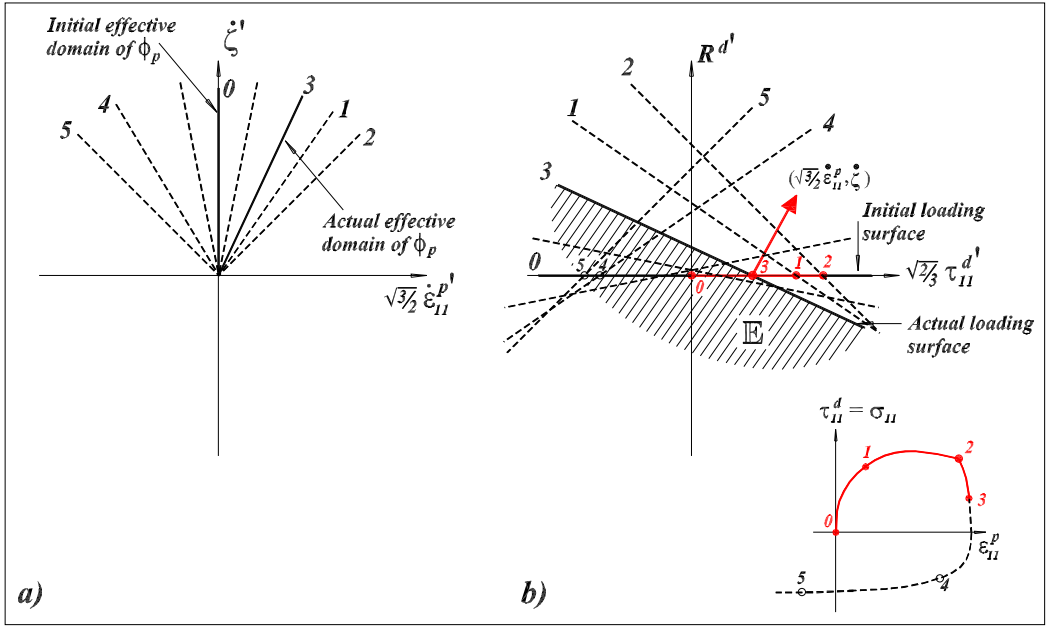}
\end{center}
\caption{Endochronic model with damage in the uniaxial stress regime. (a)
Several configurations of the effective domain $\mathbb{D}$ of the
pseudo-potential $\protect\phi_p$. (b) Corresponding configurations of the
loading domain $\mathbb{E}$, associated with the loading function $f_p$. The
actual stress point always occurs at $R^{d^{\prime}}=R^d=0$ and, at this
point, the flows $\dot{\mbox{\boldmath $\varepsilon$}}^{p}$ and $\dot{%
\protect\zeta}$ are defined by the normality condition.}
\label{Fig_domains}
\end{figure}

For the endochronic models with damage, the plasticity pseudo-potential is
defined as follows:
\begin{equation}
\phi _{p}\left( \mbox{\boldmath $\dot{\varepsilon}$}^{p^{\prime }},\dot{\zeta%
}^{\prime };\mathbf{v};\mbox{\boldmath$\rho$ }\right) =\left( 1-D\right)
\frac{\left\Vert dev\left( \mathbf{C:}\left(
\mbox{\boldmath $\varepsilon
-\varepsilon$}^{p}\right) \right) \right\Vert ^{2}}{2G/\beta }\frac{\dot{%
\zeta}^{\prime }}{g\left( \mathbf{v},\mbox{\boldmath$\rho$
}\right) }+\mathbb{I}_{\mathbb{D}}\left(
\mbox{\boldmath
$\dot{\varepsilon}$}^{p^{\prime }},\dot{\zeta}^{\prime };\mathbf{v};%
\mbox{\boldmath$\rho$ }\right)  \label{phi p}
\end{equation}%
where $\mathbb{I}_{\mathbb{D}}$ is the indicator function of the convex set
\begin{equation}
\mathbb{D=}\left\{
\begin{array}{l}
\left( \mbox{\boldmath $\dot{\varepsilon}$}^{p^{\prime }},\dot{\zeta}%
^{\prime }\right) \text{ such that } \\
tr\left( \mbox{\boldmath $\dot{\varepsilon}$}^{p^{\prime }}\right) =0,\text{
\ }\dot{\zeta}^{\prime }\geq 0\text{ and } \\
\mbox{\boldmath $\dot{\varepsilon}$}^{p^{\prime }}=\frac{dev\left( \mathbf{C:%
}\left( \mbox{\boldmath $\varepsilon
-\varepsilon$}^{p}\right) \right) }{2G/\beta }\frac{\dot{\zeta}^{\prime }}{%
g\left( \mathbf{v},\rho \right) }%
\end{array}%
\right\}
\end{equation}%
(see Figure \ref{Fig_domains}a). The first equality in $\mathbb{D}$ imposes
the plastic incompressibility of the flow. Moreover, since $D$ is supposed
to be less or equal to one and $g=g\left( \mathbf{v},\mbox{\boldmath$\rho$}%
\right) $, the hardening-softening function, is positive by assumption, the
second condition in $\mathbb{D}$ ensures the positivity of $\phi _{p}$.
Therefore, the standard properties of $\phi _{p}$, viz. non-negativity,
convexity and positive homogeneity of order 1, are matched. The third
condition in $\mathbb{D}$ gives the plastic flow and is consistent with (\ref%
{endoGen1}). It can be proven that when $%
\mbox{\boldmath
$\dot{\varepsilon}$}^{p^{\prime }}=\mbox{\boldmath
$\dot{\varepsilon}$}^{p}$ and $\dot{\zeta}^{\prime }=\dot{\zeta}$, i.e. when
the actual flows are considered, the first term of the sum in (\ref{phi p})
represents the rate of energy $\dot{e}_{p}$ dissipated by the plastic flow,
defined in (\ref{doubleIneq}) for the general case. Note that the
pseudo-potential has an additional dependence on the state variables \emph{%
and} on the past-history dependent parameters collected in $%
\mbox{\boldmath$\rho$}$.

The dual dissipation potential $\phi _{p}^{\ast }$ is obtained by the
Legendre-Fenchel transformation of $\phi _{p}$ \citep{Rockafellar69}. Since $%
\phi _{p}$ is positively homogeneous of order 1, then $\phi _{p}^{\ast }$ is
an indicator function:
\begin{equation}
\begin{array}{l}
\phi _{p}^{\ast }\left( \mbox{\boldmath$\tau$}^{d^{\prime }},R^{d^{\prime }};%
\mathbf{\ v};\mbox{\boldmath$\rho$}\right) =\underset{\left( \dot{\varepsilon%
}^{p^{\prime }},\dot{\zeta}^{\prime }\right) \in \mathbb{D}}{\sup }\left( %
\mbox{\boldmath$\tau$}^{d^{\prime }}:\mbox{\boldmath
$\dot{\varepsilon}$}^{p^{\prime }}+R^{d^{\prime }}\dot{\zeta}^{\prime }-\phi
_{p}\left( \mathbf{\dot{v}}^{\prime };\mathbf{v};\mbox{\boldmath$\rho$}%
\right) \right) \\
\text{ \ \ \ }=\underset{\left( \dot{\varepsilon}^{p^{\prime }},\dot{\zeta}%
^{\prime }\right) \in \mathbb{D}}{\sup }\left( dev\left( \mathbf{\tau }%
^{d^{\prime }}\right) \mathbf{:}\frac{dev\left( \mathbf{C:}\left(
\mbox{\boldmath $\varepsilon
-\varepsilon$}^{p}\right) \right) }{2G\text{ }g\left( \mathbf{v},\rho
\right) \text{ }/\beta }\dot{\zeta}^{\prime }+R^{d^{\prime }}\dot{\zeta}%
^{\prime }-\left( 1-D\right) \frac{\left\Vert dev\left( \mathbf{C:}\left(
\mbox{\boldmath
$\varepsilon -\varepsilon$}^{p}\right) \right) \right\Vert ^{2}}{2Gg\left(
\mathbf{v},\rho \right) /\beta }\dot{\zeta}^{\prime }\right) \\
\text{ \ \ \ }=\mathbb{I}_{\mathbb{E}}\left( \mbox{\boldmath$\tau$}%
^{d^{\prime }},R^{d^{\prime }};\mathbf{v};\mbox{\boldmath$\rho$}\right)%
\end{array}
\label{phi p star}
\end{equation}%
The indicator function $\mathbb{I}_{\mathbb{E}}$ is associated with the
convex set \newline
$\mathbb{E=}\left\{ \left( \mbox{\boldmath$\tau$}^{d^{\prime }},R^{d^{\prime
}}\right) \text{ such that }f_{p}\left( \mbox{\boldmath$
\tau$}^{d^{\prime }},R^{d^{\prime }};\mathbf{v};\mbox{\boldmath$\rho$}%
\right) \leq 0\right\} $ (see Figure \ref{Fig_domains}b) with
\begin{equation}
f_{p}\left( \mbox{\boldmath$\tau$}^{d^{\prime }},R^{d^{\prime }};\mathbf{v};%
\mbox{\boldmath$\rho$}\right) =dev\left( \mbox{\boldmath$\tau$}^{d^{\prime
}}\right) \mathbf{:}\frac{dev\left( \mathbf{C:}\left(
\mbox{\boldmath $\varepsilon
-\varepsilon$}^{p}\right) \right) }{2Gg\left( \mathbf{v},\mbox{\boldmath$%
\rho$}\right) /\beta }-\left( 1-D\right) \frac{\left\Vert dev\left( \mathbf{%
C:}\left( \mbox{\boldmath
$\varepsilon-\varepsilon$}^{p}\right) \right) \right\Vert ^{2}}{2Gg\left(
\mathbf{v},\mbox{\boldmath$\rho$}\right) /\beta }+R^{d^{\prime }}  \label{fp}
\end{equation}%
The function $f_{p}$ is the \emph{loading function} for an endochronic model
with plastic incompressibility and with isotropic damage. It is associated
with the loading domain $\mathbb{E}$. If the past-history parameter $%
\mbox{\boldmath$\rho$}$ is a scalar equal to $e_{p}$, the plastic dissipated
energy, then a \emph{work-hardening} behavior is defined, in the sense that
the loading function evolves with the plastic dissipated energy. A different
approach to define work-hardening plasticity models was proposed by %
\citet{Ristinma1999}.

The generalized normality conditions imposed on $\phi _{p}^{\ast}$ leads to:
\begin{equation}
\left\{
\begin{array}{l}
\mbox{\boldmath $\dot{\varepsilon}$}^{p}=\dot{\lambda}\frac{\partial
f_{p}\left( \mathbf{\tau }^{d},R^{d};\mathbf{v};\rho \right) }{\partial
\mathbf{\tau }^{d^{\prime }}}=\dot{\lambda}\frac{dev\left( \mathbf{C:}\left(
\mathbf{\varepsilon -\varepsilon }^{p}\right) \right) }{2Gg\left( \mathbf{v}%
, \rho\right) /\beta } \\
\dot{\zeta}=\dot{\lambda}\frac{\partial f_{p}\left( \mathbf{\tau }%
^{d},R^{d}; \mathbf{v};\rho \right) }{\partial R^{d^{\prime }}}=\dot{\lambda}
\\
\dot{\lambda}f_{p}\left( \mbox{\boldmath$\tau$}^{d},R^{d};\mathbf{v};
\mbox{\boldmath$\rho$
}\right) =0\text{ \ \ \ \ \ \ \ }f_{p}\left( \mbox{\boldmath$\tau$}
^{d},R^{d};\mathbf{v};\mbox{\boldmath$\rho$ }\right) \leq 0,\text{ \ \ \ \ \
\ }\dot{\lambda}\geq 0%
\end{array}
\right.
\end{equation}
where the last three inequalities are the Kuhn-Tucker conditions. The
plastic flow defined in (\ref{endoGen1}) is retrieved. Note that the
derivatives are taken with respect to the generic variables $%
\mbox{\boldmath$\tau$}^{d^{\prime }}$ and $R^{d^{\prime }}$, but they are
computed at the present state $\mbox{\boldmath$\tau$}^{d^{\prime }}=%
\mbox{\boldmath$\tau$}^{d}$ and $R^{d^{\prime }}\mathbf{=}R^{d}$. In
summary, the usual notions of plastic multiplier and loading surface have
been defined for an endochronic model with damage. This kind of
thermodynamic formulation for endochronic models is quite innovative and has
been first presented in \citet{ErlicherPoint2006}, for the case of no
damage. As was pointed out in that paper, an important property
characterizing endochronic models is the fact that at the actual state, the
loading function $f_{p}$ is always zero: for this reason, the consistency
condition $\dot{f}_{p}=0$ is always fulfilled as an identity and cannot be
used to compute the plastic multiplier $\dot{\lambda}$. This is also true in
this case, where the actual state is $\left( \mbox{\boldmath$\tau$}
^{d},R^{d}\right) =\left( \left( 1-D\right) \mathbf{C:}\left(%
\mbox{\boldmath
$\varepsilon-\varepsilon$}^{p}\right) ,0\right) .$ As a result, the
Kuhn-Tucker conditions reduce to $\dot{\lambda}=\dot{\zeta}\geq 0 $, where $%
\dot{\zeta}$ is the flow of the internal variable associated with $R^{d}$
and, using the language of the endochronic theory, is also the flow of the
intrinsic time measure; it can be freely defined, provided that it is
non-negative and that rate-independence is guaranteed. As already observed,
the standard choice is $\dot{\zeta}=\left\Vert dev\left(
\mbox{\boldmath
$\dot{\varepsilon}$}\right) \right\Vert$.

\begin{figure}[tbp]
\begin{center}
\includegraphics[width=17.78cm]{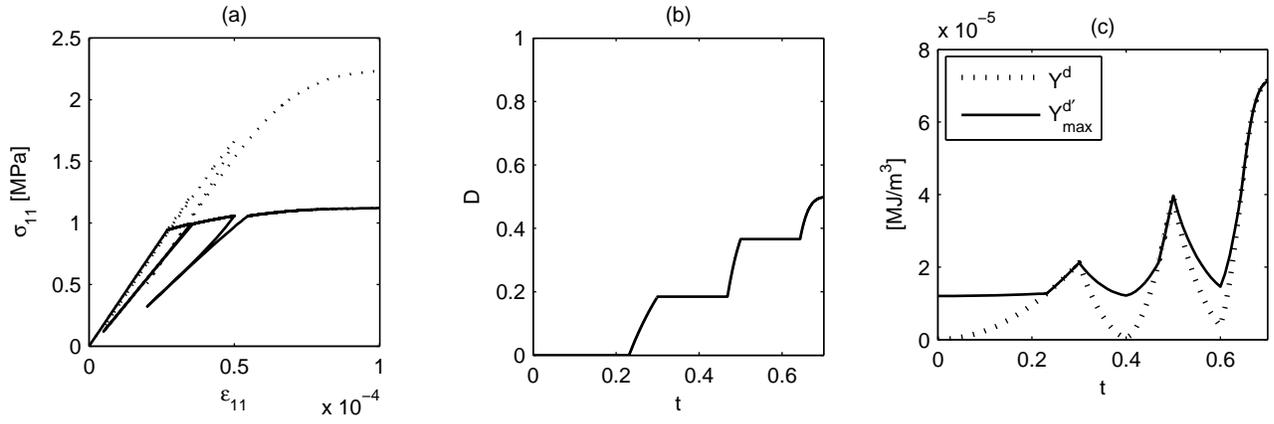}
\end{center}
\caption{Uniaxial behavior of an endochronic plasticity model with damage
(Definition 1). (a) The stress-strain evolution. The dotted line indicates
the effective stress $\protect\sigma_{11}/(1-D)$. (b) Damage evolution. (c)
See Figure \protect\ref{Fig_DamFremEl11}c.}
\label{Fig_DamFremEl-Pl11}
\end{figure}

Figure \ref{Fig_DamFremEl-Pl11} illustrates an example of uniaxial
behavior of an endochronic plasticity model with damage. The
parameters of the elastic phase and of damage (Definition 1) are the
same as those of Figure \ref{Fig_DamFremEl11}. In addition, $g=1$,
$\dot{\zeta}$ is given by ( \ref {zetanDam}) with $n=5$ , $\beta
=2834.9$ $MPa^{1-n}$ and $\gamma/\beta=-0.5$ ; as a result, $\sigma
_{u}=\left( 2G/\left(\beta +\gamma\right)\right) ^{1/n}=2.25\ast
\sqrt{2/3}=1.8371$ $MPa$, where $\sigma_{u}$ is the upper
limit of $\Vert dev\left( \mbox{\boldmath$\tau$} ^{d}/\left(1-D\right)%
\right) \Vert =\Vert dev\left( \mbox{\boldmath$\sigma$} /\left(1-D\right)%
\right) \Vert =\sqrt{2/3}\ \sigma_{11}/\left(1-D\right)$ when $g=1$.

\begin{figure}[tbp]
\begin{center}
\includegraphics[width=15cm]{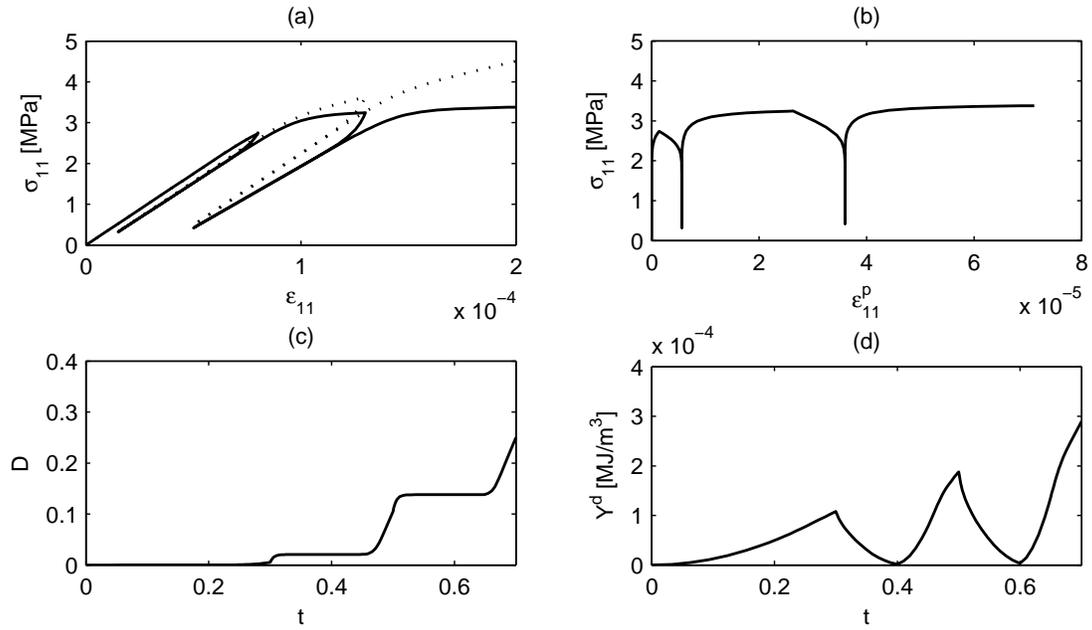}
\end{center}
\caption{Uniaxial behavior of an endochronic plasticity model with damage
(Definition 2). (a) The stress-strain evolution. The dotted line indicates
the effective stress $\protect\sigma_{11}/(1-D)$. (b) The stress as function
of the plastic strain. Observe that the plastic strain increases during
unloading phases. (c) The damage evolution, with damage slightly increasing
also during unloading phases. (d) Evolution of $Y^d$ vs. time.}
\label{Fig_EC11}
\end{figure}

In the example of Figure \ref{Fig_EC11}, the damage is defined by the rule $%
D=1-1/\left( 1+c_{\eta }e_{p}\right) $, with $c_{\eta }=1500$
$m^{3}MJ^{-1}$ (Definition 2). The parameter $c_{\eta }$ indicates
the sensitivity of damage to the energy $e_{p}$ dissipated by
plasticity. If $c_{\eta }$ is large, the damage increment at a given
$e_{p}$-value is larger than in the case of small $c_{\eta }$. The
Young modulus and the Poisson ratio are the same as in the previous
figures. The parameters defining the intrinsic time flow
(\ref{zetanDam}) are: $n=15$, $\beta =16.1846$ $MPa^{1-n}$ and
$\gamma /\beta =-0.8$ ; as a result, $\sigma _{u}=\left( 2G/\left(
\beta +\gamma \right) \right) ^{1/n}=2.25\ast \sqrt{2/3}=1.8371$
$MPa$. Moreover, the hardening function is defined as $g=\left(
1+\Vert dev\left( \mbox{\boldmath$ \varepsilon$}\left( t^{\prime
}\right) \right) \Vert _{max_{t^{\prime }\in \lbrack
0,t]}}/\varepsilon _{u}\right) ^{n}$, where $t$ is the present time
and $\varepsilon _{u}=0.0002$. Figure \ref{Fig_EC11}d depicts the
evolution of $Y^d=1/2\left( \mbox{\boldmath $\varepsilon
-\varepsilon$ }^{p}\right) :\mathbf{C:}\left( \mbox{\boldmath
$\varepsilon -\varepsilon$}^{p}\right) $, i.e. the actual value of
$Y^{d^{\prime }}$. According to (\ref{fD1}), this quantity is also
equal to $Y_{\max }^{d^{\prime }}$, which is the upper limit of
$Y^{d^{\prime }}$. The typical endochronic behavior with plastic
strains increasing during unloading phases is highlighted in Figure
\ref{Fig_EC11}b. As a result, owing to the damage rule depending on
the dissipated plastic energy, also the damage slightly increases
during the unloading phases: observe the damage evolution after
t=0.3 and t=0.5, which are the instants where unloading phases
begin.

\section{A brief discussion about stability and uniqueness}

\begin{figure}[tbp]
\begin{center}
\includegraphics[width=15cm]{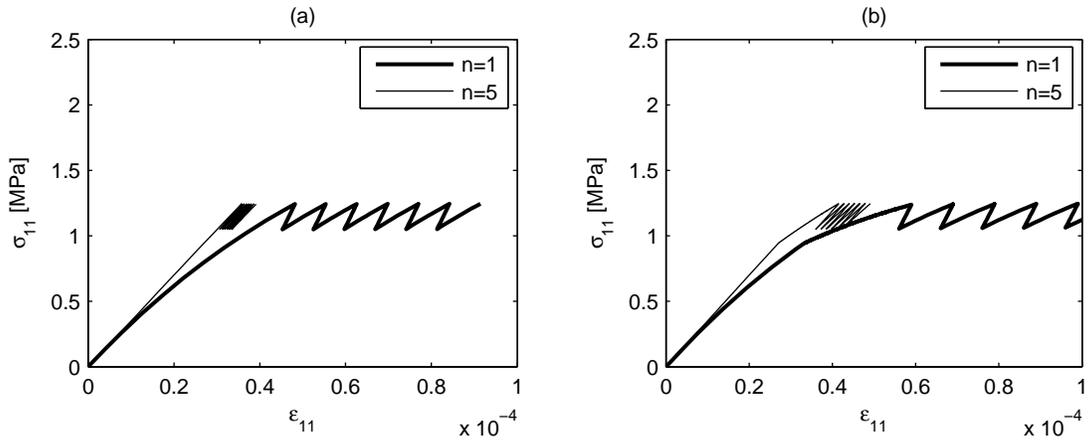} 
\end{center}
\caption{Strain accumulation, uniaxial behavior with stress varying between $%
1.05$ and $1.24$ $MPa$. (a) Endochronic model without damage, with the
intrinsic time (\protect\ref{zetan}). (b) Endochronic model with damage,
with the intrinsic time (\protect\ref{zetanDam}) and the damage evolution
given by (\protect\ref{FreDam}) and (\protect\ref{damageFlowGen}), with $%
r_0=0.000012$ and $s=6$.}
\label{Fig_Drucker}
\end{figure}

\begin{figure}[tbp]
\begin{center}
\includegraphics[width=15cm]{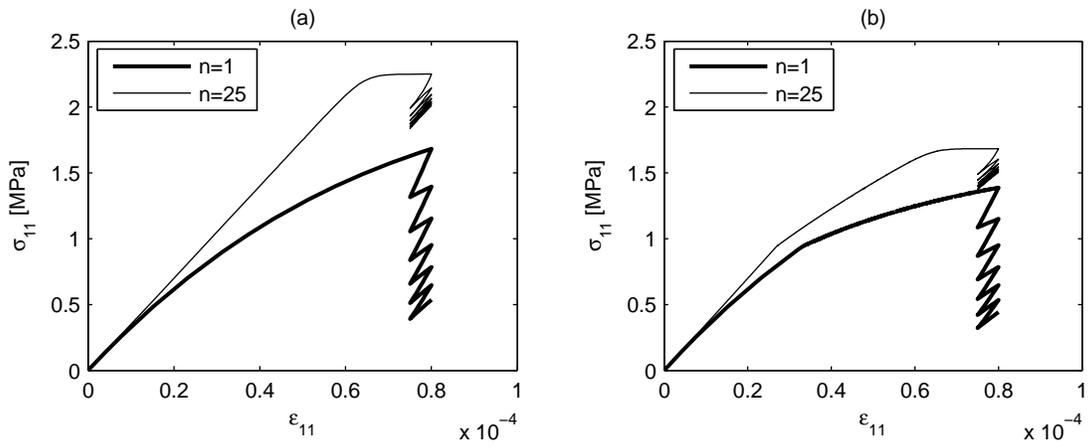} 
\end{center}
\caption{Stress relaxation, uniaxial behavior with strain varying
between 0.000075 and 0.00008. (a) Endochronic model without damage.
(b) Endochronic model with damage, with the same damage parameters
as in Figure \protect\ref {Fig_Drucker}b.} \label{Fig_Ilyushin}
\end{figure}

It is well-known that standard endochronic models violate the Drucker's
postulate and the Ilyushin's postulate, see e.g. \citep{Sandler78}. As a
result, inelastic strains may continuously increase if a cyclic stress of
constant and arbitrarily small amplitude is imposed around a given static
stress (Figure \ref{Fig_Drucker}a). Dually, a stress relaxation occurs when
cycling straining is imposed (Figure \ref{Fig_Ilyushin}a). The parameters
used for the numerical simulations of Figures \ref{Fig_Drucker} and \ref%
{Fig_Ilyushin} are: $E=35000$ $MPa$, $\nu=0.18$, $g=1$, $\gamma/\beta=-0.5$,
while $\beta$ has a value such that $\left( 2G/\left(\beta
+\gamma\right)\right) ^{1/n}=1.8371$ $MPa$, for the given $n$ values used in
the figures. The strain accumulation entails a violation of a Lyapunov-type
stability condition. For this reason, endochronic theory have been
repeatedly criticized in the past years. However, \citet[p.705]{Bazant78}
showed that endochronic models do fulfil some weaker physically motivated
stability conditions. Moreover, there are materials that are stable in the
Drucker's sense and others that are not. Hence, for these materials, a
proper model cannot fulfil the postulate of Drucker. All the aspects
concerning this subject have been explored in detail in the previously cited
references \citep{Sandler78,Bazant78} for endochronic models without damage.
A detailed analysis for the case of models with damage would deserve further
studies, but this is beyond the purposes of this paper. Figures \ref%
{Fig_Drucker}a and \ref{Fig_Ilyushin}a simply show the influence of the
parameter $n$ on the strain accumulation and the stress relaxation for an
endochronic model without damage. When $n$ tends to infinity, a plastic
behavior of Prandtl-Reuss type is retrieved, where neither strain
accumulation nor stress relaxation occur. Figures \ref{Fig_Drucker}b and \ref%
{Fig_Ilyushin}b concern models with damage.

Another important topic concerning plasticity and/or damage models is the
loss of uniqueness due to strain-softening; see e.g. \citep{JirasekBazant02}%
. An exhaustive treatment of this subject for endochronic models with damage
requires further analyses. However, for illustrative purposes, a simple
analytical study of a uniaxial model is presented hereafter. Let $\sigma $, $%
\varepsilon $ and $\varepsilon ^{p}$ be the stress, the total strain and the
plastic strain in the axial direction, respectively. Then, the uniaxial
behavior can be represented by the following law: $\sigma =\left( 1-D\right)
E\left( \varepsilon -\varepsilon ^{p}\right) =\left( 1-D\right) E\varepsilon
^{e}$, where $E$ is the Young modulus. The incremental form reads
\begin{equation}
d\sigma =\left( 1-D\right) E\left( d\varepsilon -d\varepsilon ^{p}\right)
-\sigma \frac{dD}{1-D}=\left( 1-D\right) Ed\varepsilon -\beta \sigma \frac{%
d\zeta }{g}-\sigma \frac{dD}{1-D}  \label{dsig0}
\end{equation}%
where the intrinsic time increment is
\begin{equation}
d\zeta =\left( 1+\frac{\gamma }{\beta }sign\left( \sigma d\varepsilon
\right) \right) \left\vert \sigma \right\vert ^{n-1}\left( 1-D\right)
^{1-n}\left\vert d\varepsilon \right\vert  \label{dzeta}
\end{equation}%
and the damage increment writes
\begin{equation}
dD=H\left( f_{D}\right) \frac{dR}{d\varepsilon ^{e}}d\varepsilon ^{e}\frac{%
1-D}{s\text{ }R\left( \varepsilon ^{e}\right) }  \label{dD}
\end{equation}%
with $R\left( \varepsilon ^{e}\right) =E\left\langle \varepsilon
^{e}\right\rangle ^{2}/2$ and $f_{D}=\left( 1-D\right) ^{s}R\left(
\varepsilon ^{e}\right) -r_{0}\leq 0$. Assume $\sigma >0$ and $d\varepsilon
>0$ (loading); the case $\sigma <0$, $d\varepsilon <0$ is analogous. Then,
the condition to avoid strain-softening is
\begin{equation}
d\sigma \geq 0  \label{dsig>0}
\end{equation}%
The generic damage increment when $f_{D}=0$ is given by
\begin{equation*}
dD=\frac{2}{s}\frac{1-D}{\varepsilon ^{e}}d\varepsilon ^{e}=\frac{2}{s}E%
\frac{\left( 1-D\right) ^{2}}{\sigma }\left( d\varepsilon -d\varepsilon
^{p}\right)
\end{equation*}%
Moreover, from (\ref{dsig0}) one has $d\varepsilon ^{p}=\beta \sigma d\zeta /%
\left[ E\left( 1-D\right) g\right] .$ Hence, the condition (\ref{dsig>0})
assumes the following form%
\begin{equation}
\left( \left( 1-D\right) Ed\varepsilon -\beta \sigma \frac{d\zeta }{g}
\right) \left( 1-\frac{2}{s}\right) \geq 0  \label{productSoft}
\end{equation}%
The first factor is always positive provided that $g\geq 1$. This can be
proven using the definition of $d\zeta $ given in (\ref{dzeta}) with $\sigma
>0,d\varepsilon >0$ and observing that the non-negativity of the first
factor in (\ref{productSoft}) is equivalent to the condition $\sigma /\left(
1-D\right) \leq \left( E/\left( \beta +\gamma \right) \right) ^{1/n}\left(
g\right) ^{1/n}=\sigma _{y}\left( g\right) ^{1/n}$, stating that the
effective stress is always less or equal than the bounding axial stress $%
\sigma _{y}$, modified by the hardening function $g$. If $g\geq 1$, this
inequality is always strictly fulfilled. Hence, strain-softening can be
avoided if $s\geq 2$ 
. The same result can be obtained using the tensor expressions (\ref%
{endoGen1}), (\ref{zetanDam}), (\ref{FreDam}) and (\ref{damageFlowGen}) and
imposing that all the stress components are zero except $\sigma
_{11}:=\sigma $. This proof is omitted for brevity. The same condition on $s$
has been found for the case of elasticity with damage \citep{Nedjar2001}.
Note that $g<1$ induces strain-softening also when there is no damage. The
analysis of the unloading case is not necessary, since at a given
stress-strain state with $\sigma \neq 0$, the unloading stiffness is always
greater than the loading one. 
A more complex analysis, not considered here, is needed for the multi-axial
case, where the fourth-order tensor of tangential moduli for the endochronic
model with damage should be computed. If strain-softening is avoided, the
uniaxial behavior in what concerns the strain accumulation and the stress
relaxation is analogous to that of standard endochronic models.

\section{Conclusions}

An extended endochronic theory with a scalar damage variable was developed,
based on the postulate of strain equivalence and by using pseudo-potentials
depending on state variables and on parameters related to the past history
of the material. The relevant loading surfaces, for damage and for
plasticity, were defined. Two different damage pseudo-potentials were
discussed and a formalization of the conditions on state variables affecting
the definition of damage was provided, by an additional indicator function
in the Helmholtz free energy. In a companion paper \citep{ErlicherBursi2007}
, a link between this extended endochronic theory and the Bouc-Wen type
models with both strength and stiffness degradation is established. This
will permit to prove the thermodynamic admissibility of these Bouc-Wen
models and to highlight a constraint for the relevant stiffness degradation
rules.

\section{Appendix: Notations}

\textit{The following symbols are used in this paper:}

$\mathbf{C}=$ fourth-order elasticity tensor

$D=$ internal variable associated with isotropic damage

$e_{D}=$ energy per unit volume dissipated through damage

$e_{p}=$ energy per unit volume dissipated through plasticity

$f_{D}=$ loading function for damage

$f_{p}=$ loading function for plasticity

$G=$ shear modulus

$g=$ hardening-softening function

$H(\cdot)=$ Heaviside function

$\mathbf{I}=$ fourth-order identity tensor

$\mathbb{I}_{\mathbb{H}}=$ indicator function of the set $\mathbb{H}$

$K=$ bulk modulus

$\mathbf{q}^{d}=$ dissipative thermodynamic forces vector

$\mathbf{q}^{nd}=$ non-dissipative thermodynamic forces vector

$R^{d}=$ dissipative part of the thermodynamic force introducing isotropic
hardening(softening)

$R^{nd}=$ non-dissipative part of the thermodynamic force introducing
isotropic hardening(softening)

$\mathbf{v}=$ state variables vector

$Y^{d}=$ dissipative part of the thermodynamic force dual to the damage
variable

$Y^{nd}=$ non-dissipative part of the thermodynamic force dual to the damage
variable

$\mathbf{z}=$ \emph{hysteretic} part of the stress tensor

$\beta =$ coefficient defining the plastic flow of Endochronic models

$\gamma =$ coefficient defining the plastic flow of Endochronic models

$\mbox{\boldmath $\varepsilon$}=$ total small strain tensor

$\mbox{\boldmath $\varepsilon$ }^{p}=$ plastic small strain tensor

$\zeta =$ intrinsic time \emph{measure} for Endochronic models. Moreover, it
is the internal variable associated with isotropic hardening/softening of
Endochronic models

$\vartheta =$ intrinsic time \emph{scale} for Endochronic models

$\dot{\lambda} =$ plastic multiplier

$\dot{\lambda_{D}}=$ damage multiplier

$\mu =$ hereditary kernel

$\mbox{\boldmath $\rho$}=$ history-dependent parameters vector

$\mbox{\boldmath $\sigma$}=$ Cauchy stress tensor

$\mbox{\boldmath $\sigma$}^{d}=$ dissipative part of the Cauchy stress tensor

$\mbox{\boldmath $\sigma$}^{nd}=$ non-dissipative part of the Cauchy stress
tensor

$\mbox{\boldmath $\tau$}^{d}=$ dissipative part of the thermodynamic force
dual to the plastic strain tensor

$\mbox{\boldmath $\tau$}^{nd}=$ non-dissipative part of the thermodynamic
force dual to the plastic strain tensor

$\Phi _{1}=$ mechanical or intrinsic dissipation

$\phi =$ pseudo-potential or dissipation potential for plasticity

$\phi ^{\ast }=$ dual pseudo-potential for plasticity

$\phi _{D}=$ pseudo-potential or dissipation potential for damage

$\phi _{D}^{\ast }=$ dual pseudo-potential for damage

$\Psi =$ Helmholtz free energy volume density

$\mathbf{1}=$ second-order identity tensor

$\langle\rangle$=McCauley brackets

\clearpage \listoffigures \clearpage

\clearpage

\clearpage

\clearpage

\clearpage

\clearpage

\clearpage

\clearpage

\clearpage

\end{document}